\documentclass[aps,prd,reprint,onecolumn,superscriptaddress,showkeys]{revtex4-2}
\usepackage{graphicx}
\usepackage{amsmath}
\usepackage[dvipsnames]{xcolor}
\usepackage{upgreek}
\usepackage{caption}
\usepackage{natbib}
\pdfoutput=1
\usepackage{hyperref}
\hypersetup{nolinks=true}

\begin{document}
\title{Simulating noise on a quantum processor: interactions between a qubit and resonant two-level system bath}

\author{Yujin Cho}
\email{cho25@llnl.gov}
\affiliation{Lawrence Livermore National Laboratory, Livermore, California 94550, USA}
\author{Dipti Jasrasaria}
\affiliation{Department of Chemistry, University of California, Berkeley, California 94720, USA}
\author{Keith G. Ray}
\affiliation{Lawrence Livermore National Laboratory, Livermore, California 94550, USA}
\author{Daniel M. Tennant}
\thanks{Current address: Rigetti Computing, 775 Heinz Ave., Berkeley, California 94710, USA}
\affiliation{Lawrence Livermore National Laboratory, Livermore, California 94550, USA}
\author{Vincenzo Lordi}
\affiliation{Lawrence Livermore National Laboratory, Livermore, California 94550, USA}
\author{Jonathan L DuBois}
\affiliation{Lawrence Livermore National Laboratory, Livermore, California 94550, USA}
\author{Yaniv J. Rosen}
\affiliation{Lawrence Livermore National Laboratory, Livermore, California 94550, USA}

\begin{abstract}

Material defects fundamentally limit the coherence times of superconducting qubits, and manufacturing completely defect-free devices is not yet possible. Therefore, understanding the interactions between defects and a qubit in a real quantum processor design is essential. We build a model that incorporates the standard tunneling model, the electric field distributions in the qubit, and open quantum system dynamics, and draws from the current understanding of two-level system (TLS) theory.
Specifically, we start with one million TLSs distributed on the surface of a qubit and pick the 200 systems that are most strongly coupled to the qubit. We then perform a full Lindbladian simulation that explicitly includes the coherent coupling between the qubit and the TLS bath to model the time dependent density matrix of resonant TLS defects and the qubit. We find that the 200 most strongly coupled TLSs can accurately describe the qubit energy relaxation time. This work confirms that resonant TLSs located in areas where the electric field is strong can significantly affect the qubit relaxation time, even if they are located far from the Josephson junction. Similarly, a strongly-coupled resonant TLS located in the Josephson junction does not guarantee a reduced qubit relaxation time if a more strongly coupled TLS is far from the Josephson junction. 
In addition to the coupling strengths between TLSs and the qubit, the model predicts that the geometry of the device and the TLS relaxation time play a significant role in qubit dynamics. Our work can provide guidance for future quantum processor designs with improved qubit coherence times.

\end{abstract}

\maketitle

\section{Introduction}

Due to the flexibility in design and compatibility with existing CMOS fabrication technologies \cite{devoret_implementing_2004, kjaergaard_superconducting_2020}, superconducting qubits are a promising platform for complicated calculations, including simulating quantum phenomena found in nature \cite{wendin_quantum_2017} and potentially performing large scale calculations that are otherwise impractical with classical computers \cite{shor_polynomial-time_1997}. Recently, small scale simulations in quantum chemistry and nuclear physics have demonstrated these possibilities \cite{google_ai_quantum_and_collaborators_hartree-fock_2020,shi_simulating_2021}. To further advance such capabilities, long coherence times of qubits are essential. While the coherence properties of superconducting qubits have advanced significantly over the last two decades, the qubit performance is fundamentally limited by material defects in superconducting devices \cite{muller_towards_2019}.

Defects induce decoherence in superconducting qubits through various processes. Resonant defects, which have energies that are close to the qubit energy, cause energy relaxation \cite{wang_surface_2015, muller_towards_2019, de_Graaf_two-level_2020}, while non-resonant defects contribute to dephasing \cite{faoro_microscopic_2008, bergli_decoherence_2009, schlor_correlating_2019}. In an effort to better understand these processes, several papers have simulated the interaction between a single resonant defect and a qubit or superconducting resonator \cite{ku_decoherence_2005, bhattacharya_jaynes-cummings_2011, bhattacharya_understanding_2012, rosen_protecting_2019}. Although these results introduce a useful framework for understanding qubit-defect interactions, they do not accurately describe the qubit dynamics of a realistic device with a non-uniform electric field distribution that interacts with many defects \cite{martinis_decoherence_2005, muller_towards_2019}. Therefore, an accurate description of the interaction between a qubit and the material defects on a real quantum processor design is essential to the development of qubit designs that are less sensitive to noise. 

Our work directly simulates the interactions between a qubit and an ensemble of resonant defects, called two-level system (TLS) defects \cite{muller_towards_2019, phillips_two-level_1987}, using the full Lindblad master equation \cite{manzano_short_2020}. We explicitly model the ensemble of TLSs as individual quantum states with their own decay channels to the environment. 
In addition, we consider the non-uniform electric field distribution representative of real devices, which imposes a range of coupling strengths for the TLSs distributed across the surfaces. 

To model the interactions between a qubit and many defects, we utilize the standard tunneling model (STM) \cite{lisenfeld_decoherence_2016, muller_towards_2019}. This model treats the defects as atoms tunneling between two local minima in a double-well potential \cite{phillips_two-level_1987, anderson_anomalous_1972}, approximated by TLSs. We use the parameters drawn from this model to distribute one million TLSs on a realistic and widely used quantum processor design and select the 200 most strongly coupled TLSs to simulate the qubit dynamics. This study shows that the position of the strongest coupled TLS, not just the coupling strength, informs the qubit dynamics. In addition, at least 150 TLSs are necessary to accurately describe the qubit dynamics in a general case. Lastly, using experimental observations, we explore and put bounds on TLS relaxation times, which cannot be predicted by the STM. The method we developed in this work bridges the gap between an ensemble picture of open quantum systems and a single TLS model. 

\section{Methods}

\begin{center}
\begin{figure}[t]
\includegraphics[scale=1.0]{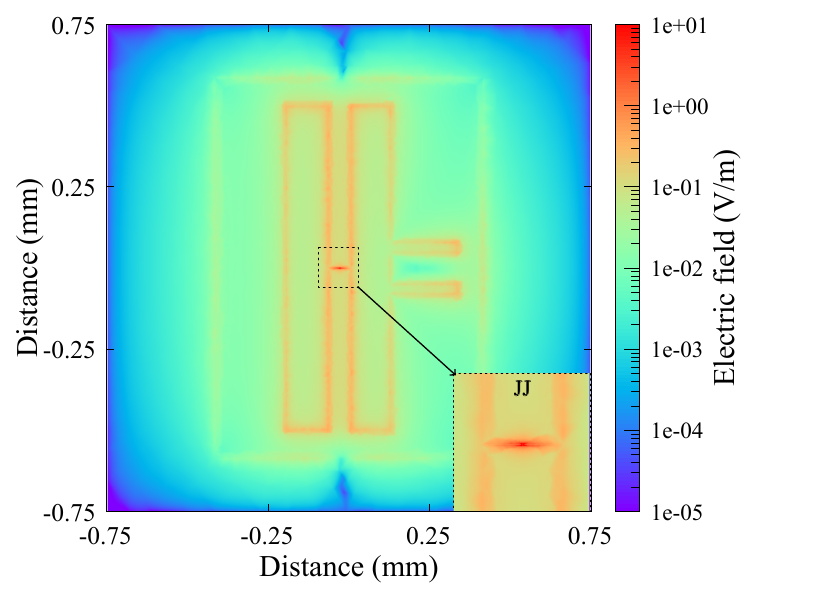}
\caption{The electric field distribution of the superconducting qubit has a large amplitude, reaching $10\,\mathrm{V/m}$ at the Josephson junction in the center of the panel, and $\sim1\,\mathrm{V/m}$ around the edges of the capacitance pads. The Josephson junction size is $250\times350\,\mathrm{nm^2}$.}
\label{fig:Efield}
\end{figure}
\end{center}

To simulate qubit-TLS dynamics on a realistic quantum processor design, we calculate the electric field distributions using Ansys HFSS on a transmon qubit \cite{place_new_2021} with a single Josephson junction (JJ). The electric field strength was initially evaluated with the structure oscillating at $1\,\textrm{Joule}$ and was later scaled to a single photon at the frequency of $5\,\mathrm{GHz}$. Figure~\ref{fig:Efield} shows that the electric field in the JJ, which is located at the center of the transmon, is at least one order of magnitude stronger than in other parts of the chip, $\sim10\,\mathrm{V/m}$. The edges of the capacitor pad structures also have relatively strong electric fields (orange color), around $1\,\mathrm{V/m}$.

Using these realistic electric field distributions, we randomly distributed one million TLSs across the device, in the first $3\,\mathrm{nm}$ of the substrate, where the dominant surface loss is considered to occur \cite{wenner_surface_2011}. The dipole moment of each TLS defect was initially fixed at 3 Debye ($\sim0.6\,\mathrm{e{\AA}}$) \cite{hung_probing_2022}.
To account for the resonantly coupled TLSs to the qubit, which has typical coherence times longer than $0.1\,\mathrm{\upmu s}$, we chose the TLS frequencies to be within $10\,\mathrm{MHz}$ of the qubit frequency at $5\,\mathrm{GHz}$.

In this work, we assume that the qubit energy relaxes only via resonant TLS defects, and we ignore the interactions between the qubit and off-resonant TLSs, quasiparticles, and other environmental modes. Additionally, we initialize the system with a single photon such that the system can only lose energy without energy gain. These two assumptions allow us to simplify the Hamiltonian. The total Hamiltonian in this system can be written as a sum of four terms: 1) the qubit Hamiltonian, $H_\textrm{q}$, 2) the TLS Hamiltonian, $H_\textrm{TLS}$, 3) the qubit-TLS interaction Hamiltonian, $H_\textrm{q-TLS}$, and 4) the TLS-TLS interaction Hamiltonian, $H_\textrm{TLS-TLS}$.
\begin{equation}
H = H_\textrm{q} + H_\textrm{TLS} + H_\textrm{q-TLS} + H_\textrm{TLS-TLS}\,.
\label{eqn:Hamiltonian}
\end{equation}
The qubit Hamiltonian, $H_\textrm{q}$, is defined as $H_\textrm{q} = E_q\,\sigma_z^q/2$ in its eigenbasis, where the qubit energy is $E_q = \hbar\omega_q$ and we set $\omega_q=2\pi\cdot5\,\mathrm{GHz}$.

In the standard tunneling model, a tunneling atom can redistribute charge in the material, leading to a dipole moment, $\vec{p}$. A TLS can be described using a double-well potential that has an energy difference between the two energy levels, $\Delta$, and a tunneling energy, $\Delta_0$ \cite{phillips_two-level_1987}.
In this picture, a single TLS Hamiltonian can be written as $\tilde{H}_\textrm{TLS}=\Delta\tilde{\sigma}_z^\textrm{TLS}/2 + \Delta_0 \tilde{\sigma}_x^\textrm{TLS}/2$. When $\tilde{H}_\textrm{TLS}$ is transformed to its eigenbasis, $H_\textrm{TLS}=E_{TLS} \sigma_z^{TLS} /2$, where $E_{TLS}=\sqrt{(\Delta_{TLS})^2+(\Delta_{0,TLS})^2}$ is the TLS energy. 
The qubit-TLS interaction Hamiltonian can be written as $H_\textrm{q-TLS}=\sum_i \hbar g_i\sigma_x^q \tilde{\sigma}_x^{TLS}$, where $g_i=\vec{p}_i\cdot \vec{E_i^q}/\hbar$ is the coupling strength between the $i$-th TLS and the qubit, for the TLS dipole moment, $\vec{p}_i$, and the local electric field generated by the qubit oscillation at the position of the TLS, $\vec{E_i^q}$. In TLS eigenbasis, 
\begin{equation}
\begin{split}
H_\textrm{q-TLS}&=\sum_i \hbar g_i\sigma_x^q \Big(\frac{\Delta_{0,TLS}^i}{E_{TLS}^i}\sigma_x^{TLS}+\frac{\Delta_{TLS}^i}{E_{TLS}^i}\sigma_z^{TLS}\Big)\\
&\approx\sum_i \hbar g_i \frac{\Delta_{0,TLS}^i}{E_{TLS}^i}\sigma_x^q \sigma_x^{TLS}\,.
\end{split}
\label{eqn:qTLS_ham}
\end{equation}
Because $\sigma_x^q \sigma_z^{TLS}$ coupling is much weaker than $\sigma_x^q \sigma_x^{TLS}$ coupling, the $\sigma_x^q \sigma_z^{TLS}$ term can be ignored.
Then we apply rotating wave approximation to $H_{q-TLS}$:
\begin{equation}
H_\textrm{q-TLS}\approx\sum_i \hbar \Omega_i (\sigma_-^q\sigma_+^{TLS} + \sigma_+^q\sigma_-^{TLS})\,.
\label{eqn:qTLS_ham_rwa}
\end{equation}
In Eq. (\ref{eqn:qTLS_ham_rwa}), we define the effective coupling strength $\Omega_i$ as: 
\begin{equation}
\Omega_i = g_i\frac{\Delta_{0}^i}{E^i_{\textrm{TLS}}}\,\,.
\label{eqn:rabi}
\end{equation}

$H_\textrm{TLS-TLS}$ represents the interaction between a pair of TLSs. In the TLS eigenbasis, the dipole-like interaction between TLSs is \cite{agarwal_polaronic_2013, black_spectral_1977}
\begin{equation}
H_\textrm{TLS-TLS}=\frac{C_{ij}(\hat{r_{ij}})}{4} \frac{1}{r_{ij}^3}\Big(\frac{\Delta_{TLS}^i}{E^i_{TLS}} \sigma^{TLS,i}_z + \frac{\Delta_{0,TLS}^i}{E^i_{TLS}} \sigma^{TLS,i}_x \Big)\Big(\frac{\Delta_{TLS}^j}{E^j_{TLS}} \sigma^{TLS,j}_z + \frac{\Delta_{0,TLS}^j}{E^j_{TLS}} \sigma^{TLS,j}_x\Big)\,,
\label{eqn:interaction}
\end{equation}
where $C_{ij}(\hat{r_{ij}})$ is a material-specific constant that depends on the vector between $i$-th and $j$-th TLS positions. 
In Eq. (5), the $\sigma_x^{TLS,i} \sigma_z^{TLS,j} $ term is much smaller than the $\sigma_x^{TLS,i}  \sigma_x^{TLS,j}$ term, and the $\sigma_z^{TLS,i} \sigma_z^{TLS,j} $ term vanishes because we only have a single photon in the system and do not have mechanisms for energy gain.
Therefore, $H_\textrm{TLS-TLS}$ simplifies to:
\begin{equation}
H_\textrm{TLS-TLS}=\frac{C_{ij}(\hat{r_{ij}})}{4} \frac{1}{r_{ij}^3} \frac{\Delta_{0,TLS}^i}{E^i_{TLS}} \frac{\Delta_{0,TLS}^j}{E^j_{TLS}} \sigma^{TLS,i}_x \sigma^{TLS,j}_x\,.
\end{equation}
Instead of using $C_{ij}(\hat{r_{ij}})$ for $i$- and $j$-th TLSs, we used the root-mean-square (rms) angular averaged value, $C_{rms}=1.6\times 10^{-48}\,\mathrm{J\cdot m^3}$ \cite{black_spectral_1977}. We also chose the same dipole moment, $|\vec{p}|$, for all TLSs between 0.1 and 4.0 Debye \cite{lisenfeld_electric_2019, hung_probing_2022}, and $\vec{E}_i^q$ was sampled at the location of each TLS from the electric field simulation, as shown in Fig. \ref{fig:Efield}. The angle between $\vec{p}$ and $\vec{E}_i^q$ was randomly selected between 0 and $\pi$.

The tunneling energies of TLSs, $\Delta_{0,TLS}$, are randomly sampled from a distribution, as follows. 
In the standard tunneling model, the energy density of the TLS ensemble is independent of $\Delta$ and is equal to $P_0/\Delta_0$ \cite{phillips_tunneling_1972, phillips_two-level_1987}.
For the total number of the TLSs, $N$, in a volume $V$,
\begin{equation}
N = \iint \frac{P_0 V}{\Delta_{0,TLS}}\,d\Delta_{0,TLS}\,d\Delta_{TLS}\,\,, 
\label{eqn:d}
\end{equation}
where $P_0$ is the uniform TLS energy density given by $10^{44}\,\mathrm{/J\cdot m^3}$ \cite{khalil_landau-zener_2014}. 

In Eq.(\ref{eqn:d}), as $\Delta_0$ goes to zero, $N$ goes to infinity. Thus, to prevent the divergence of $N$, we excluded TLSs with very low $\Delta_0$. In the regime of very small $\Delta_0$, the coupling strength between the TLS and the qubit is weak and can be safely ignored.

In polar coordinates, Eq.(\ref{eqn:d}) can be expressed as
\begin{equation}
N=\int_d^{\pi/2}\,\int_{E_q-\delta}^{E_q+\delta}\frac{P_0 V}{\mathrm{sin}\theta}dE\,d\theta\,,
\label{eqn:d1}
\end{equation}
where $E_q$ is the qubit energy, $\delta$ is the bandwidth of the TLS energies, $10\,\mathrm{MHz}$, $\mathrm{sin}\theta=\Delta_0/E$, and $d$ is a dimensionless parameter that sets the lower bound of $\theta$. $\Delta_0$ of each TLS was then randomly selected above $d$. For $N=10^6$ and $V=1.5\times1.5\times3\times10^{-15}\,\mathrm{m^3}$, $d=2\exp(-10^4)$, which is vanishingly small and should encompass all relevant TLSs.

As we will show, the energy relaxation dynamics of the qubit are sensitive to the number and positions of strongly coupled TLSs. Therefore, we model the qubit dynamics due to the qubit-TLS interactions in a realistic qubit design, while allowing the TLSs to couple to an open quantum environment. We restrict the system to one photon, which mimics real world operation and allows us to reduce the size of the Hilbert space.

The qubit-TLS system can be described by a time-dependent density matrix, $\rho(t)$, whose dynamics can be described with the Lindblad master equation \cite{manzano_short_2020}:
\begin{equation}
\dot{\rho}(t)=-\frac{i}{\hbar}\left[ H,\rho(t)\right] + \sum_i \frac{1}{2}\left( 2 C_i\rho C_i^{\dagger}-\{ C_i^{\dagger}C_i, \rho \}\right)\,.
\label{eqn:Lindblad}
\end{equation}
The first term in Eq. (\ref{eqn:Lindblad}) describes the unitary dynamics between the qubit and the TLSs, where $H$ is given in Eq. (\ref{eqn:Hamiltonian}). The second term describes the TLSs emitting phonons to the environment, which is the only relaxation channel to the environment in the model.
The collapse operator, $C_i=a/\sqrt{T_1^{\textrm{TLS},i}}$, represents phonon emission from the $i$-th TLS at a rate $1/T_1^{\textrm{TLS},i}$, where $a$ is the annihilation operator and $T_1^{\textrm{TLS},i}$ is the energy relaxation time of the $i$-th TLS. Phonon emission is the primary relaxation channel for TLSs \cite{muller_towards_2019}. In the standard tunneling model, the TLS energy decay rate depends on $\Delta_0$ of the TLS, which can be derived through a perturbative interaction between the TLS and the lattice strain field \cite{phillips_two-level_1987} and expressed as:
\begin{equation}
T_1^{\textrm{TLS},i}=T_\textrm{1,min}^\textrm{TLS} / (\Delta_{0,norm}^i)^2\,.
\label{eqn:T1min}
\end{equation}
A dimensionless value, $\Delta_{0,norm}^i$, is $\Delta_0$ of the $i$-th TLS that is normalized to its energy, and $T_\textrm{1,min}^\textrm{TLS}$ is a constant that we treat as an adjustable parameter that will be discussed later.
We use the QuTiP python package \cite{johansson_qutip_2012, johansson_qutip_2013} to solve the Lindblad master equation dynamics. 
The qubit energy relaxation time, $T_1^q$, is obtained by tracking the diagonal component in the density matrix at each time step with the initial state of the qubit at $|1\rangle$. Similarly, the qubit pure dephasing time, $T_2^q$, is acquired from the change in the off-diagonal component, with the initial state at $(|0\rangle+|1\rangle)/\sqrt2$. The simulations were performed at the Livermore High Performance Computing facility.

\section{Results}

\begin{center}
\begin{figure}[t]
\includegraphics[scale=1.0]{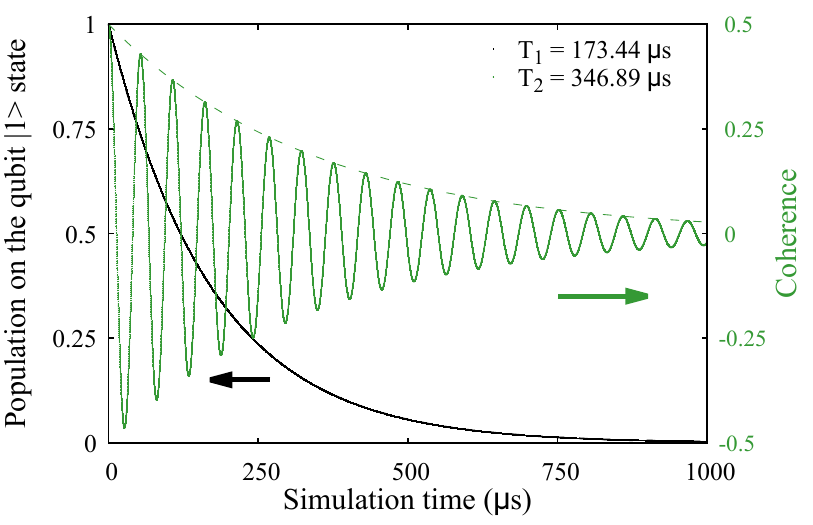}
\caption{(b) Example of simulated qubit energy relaxation (left axis, black) and dephasing (right axis, green) in this geometry. $T_1$ shows good agreement with typical experimental data, meaning our model closely describes the dynamics of a real device.}
\label{fig:T1T2}
\end{figure}
\end{center}

Figure~\ref{fig:T1T2} shows the $T_1^q$ (black, left axis) and $T_2^q$ (green, right axis) curves of the qubit with a representative distribution of TLS defects on the qubit design in Fig.~\ref{fig:Efield}. The qubit energy relaxation curve is consistent with the typical response of a qubit \cite{place_new_2021,tennant_low-frequency_2022}. As we have neglected additional dephasing channels in our simulation, the dephasing time is limited by the energy relaxation and, therefore, is equal to twice the qubit $T_1$ time. 
\begin{center}
\begin{figure}[t]
\includegraphics[scale=1.0]{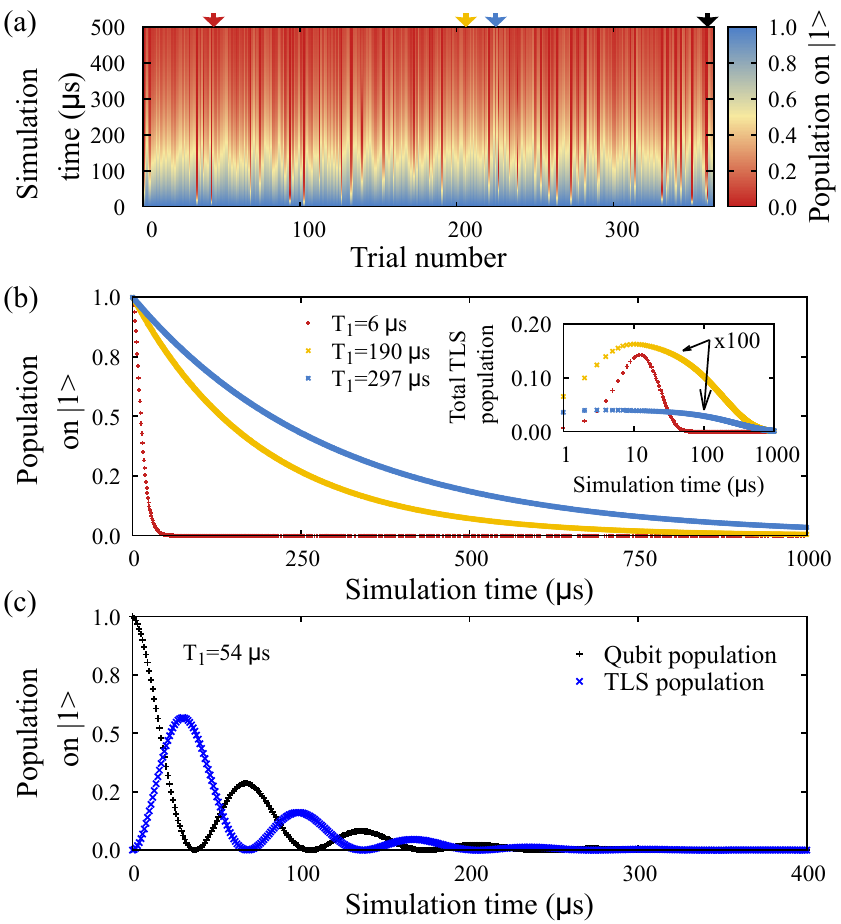}
\caption{(a) Simulated time-dependent qubit population from the excited to ground state for 364 trial cases. The color indicates when the qubit is in its excited (blue) and ground state (red). (b) Linecuts of panel (a) in the three cases indicated by the colored arrows in panel (a). They show the simulated energy relaxation curves when $T_1^q$ is the longest ($296.81\,\upmu\mathrm{s}$, blue), average value ($189.96\,\upmu\mathrm{s}$, yellow), and shortest ($6.33\,\upmu \mathrm{s}$, red). The inset shows the corresponding change in the total population of TLS excited states as a function of time. (c) When $\Delta_0$ of the strongly coupled TLS is small and $\vec{p}\cdot\vec{E}$ is large, a case indicated by the black arrow in panel (a), we observe population exchange between the qubit excited state (black) and the collective TLS excited state (blue).}
\label{fig:tddata}
\end{figure}
\end{center}

Experimentally, when a qubit undergoes a temperature-cycle above and below its superconducting gap, the TLS defects are redistributed. Also, different devices with the same qubit geometry could have completely different TLS distributions. To mimic such behavior, we simulated several trials, resampling the TLSs at each trial so that each new set of TLSs have different energies and coupling strengths to the qubit. Figure~\ref{fig:tddata}(a) shows the qubit energy relaxation over $500\,\upmu\mathrm{s}$ for 364 trials. The TLS dipole moments were fixed at 3 Debye and $T_\textrm{1,min}^\textrm{TLS}$ was fixed at $0.05\,\upmu\mathrm{s}$. 

Figure~\ref{fig:tddata}(b) shows three linecuts from panel (a) when the $T_1^q$ is the shortest (red), average (yellow), and the longest (blue). These three cases are also marked in panel (a) with arrows of the same colors. When $T_1^q$ is the shortest, $\sim6\,\upmu\mathrm{s}$, a portion of the population on the qubit excited state is transferred to the TLS excited states (see the inset in panel(b)) and eventually dissipates. On the other hand, when the $T_1^q$ is long, the total TLS population throughout the simulation time is two orders of magnitude lower. 

In a few cases, we discover oscillations in the population on the qubit excited state as shown in Fig.~\ref{fig:tddata}(c), also marked by the black arrow in panel (a). Such oscillations show the coherent exchange of population between the qubit and a strongly coupled TLS. In these cases, the qubit $T_1$ is defined as the decaying constant of the oscillation amplitude. 

\begin{center}
\begin{figure}[t]
\includegraphics[scale=1.0]{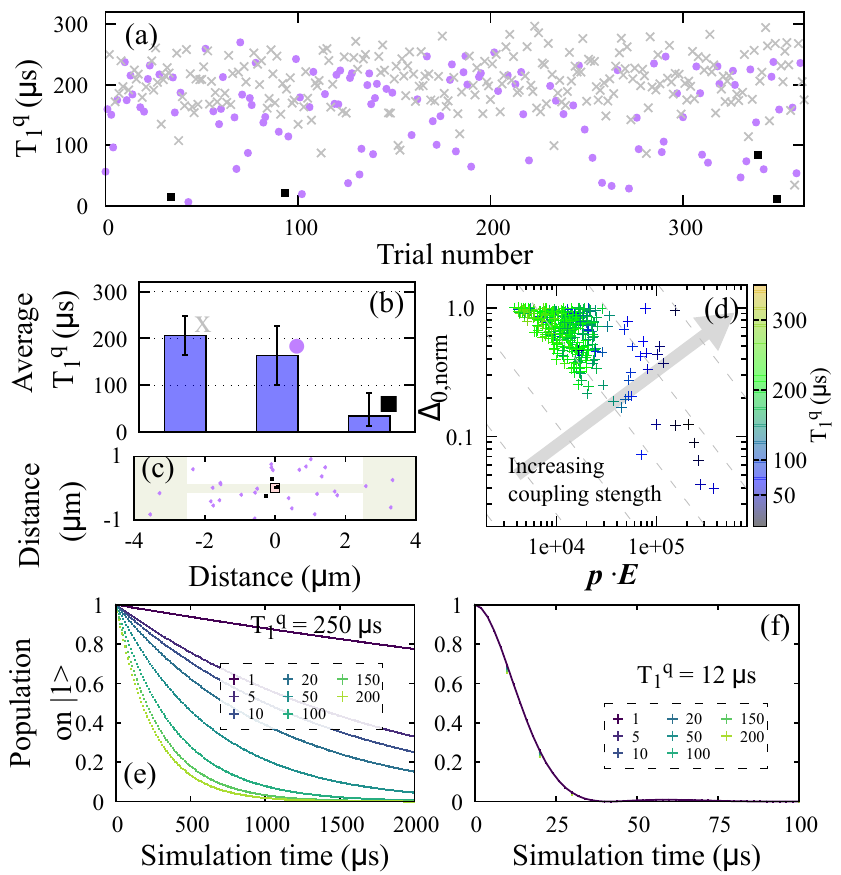}
\caption{(a) The qubit energy relaxation time, $T_1^q$, changes for different trial cases. In each case, we redistribute the TLS positions and energies. Black squares indicate that the strongest coupled TLS is located less than $0.3\,\upmu\mathrm{m}$ from the JJ, purple circles indicate that the strongest coupled TLS location is  $0.3-5\,\upmu\mathrm{m}$ from the JJ, and grey crosses are for the remaining trials. (b) Average $T_1^q$ times of grey, purple, and black points show that the $T_1^q$ is the shortest when the strongest coupled TLS is close to the JJ. (c) The position of the strongest coupled TLSs are shown on the zoomed-in transmon design. The shaded region is the thin aluminum metal and the red shaded area with solid square borderline at the center is the JJ. 
(d) Relationship between the TLS tunneling energy, $\Delta_0$, and $\vec{p}\cdot\vec{E}$ shows that the $T_1^q$ time (color bar) decreases as $\vec{p}\cdot\vec{E}$ increases. Grey dashed lines represent the equal-coupling strength lines. (e-f) Qubit dynamics with different numbers of simulated TLSs interacting with the qubit. (e) The relaxation of a long-lived qubit ($T_1^q=250\,\mathrm{\upmu s}$ with 200 TLSs) changes with the number of TLSs. (f) In the case of a short-lived qubit ($T_1^q=12\,\mathrm{\upmu s}$ with 200 TLSs), a single TLS is enough to describe the qubit dynamics.}
\label{fig:tdanalysis}
\end{figure}
\end{center}

To gain a deeper understanding of the TLS-qubit interaction, we explored the relationship between the $T_1^q$ time and the TLS parameters in the trials. First, we graph the relationship between the $T_1^q$ and the distance of the strongest coupled TLS to the qubit from the center of the JJ. 
Figure~\ref{fig:tdanalysis}(a,b) shows the $T_1^q$ times extracted from Fig.~\ref{fig:tddata}(a) colored by the most strongly coupled TLS's proximity to the JJ in the qubit. Figure \ref{fig:tdanalysis}(c) shows the qubit structure zoomed in around the JJ with the locations of the one strongest coupled TLS in each trial run. The median $T_1^q$ is around $200\,\upmu\mathrm{s}$, which is the same order of magnitude as experimentally observed qubit relaxation times \cite{tennant_low-frequency_2022, gordon_environmental_2022}. The $T_1^q$ drops as low as $6\,\upmu\mathrm{s}$ for certain TLS distributions.

When the strongest coupled TLS is within $0.3\,\upmu\mathrm{m}$ from the JJ (black squares), the $T_1^q$ drops sharply and the average $T_1^q$ is $\sim33\,\upmu\mathrm{s}$. The shortest $T_1^q$ of the black squares is $\sim12\,\upmu\mathrm{s}$ and the longest is $\sim83\,\upmu\mathrm{s}$. When the strongest coupled TLS is $0.3 - 5\,\upmu\mathrm{m}$ from the JJ (purple circles), the $T_1^q$ time depends significantly on the relative angle of the TLS dipole moment to its local electric field, which gives the largest standard deviation, $\sim63\,\upmu\mathrm{s}$, shown in panel (b). When the strongest coupled TLS is further away (grey crosses), the average $T_1^q$ is $205\pm42\,\upmu\mathrm{s}$.

In general, the electric field becomes stronger with proximity to the JJ, which results in stronger coupling strength for TLSs. Surprisingly, the coupling strength, which is the $\Delta_0$ times $\vec{p}\cdot\vec{E}$, is not a good indicator for the low $T_1^q$ times. Instead, $T_1^q$ is mostly related to only $\vec{p}\cdot\vec{E}$ of the strongest coupled TLS. 
Figure~\ref{fig:tdanalysis}(d) shows the correlation between the tunneling energy normalized to the TLS energy ($\Delta_{0, \textrm{norm}}$) and $\vec{p}\cdot\vec{E}$ of the strongest coupled TLSs. The color indicates the $T_1^q$ in each trial, and the grey dashed lines illustrate equal-coupling strength lines. This analysis shows that even if the coupling strength is the same, the $T_1^q$ time decreases with stronger $\vec{p}\cdot\vec{E}$, whereas $\Delta_0$ is weakly correlated with $T_1^q$. Moreover, if $\vec{p}\cdot\vec{E}$ of one of the most strongly coupled TLSs is high ($>10^5\,\mathrm{Hz}$), and $\Delta_0$ is low ($\Delta_\textrm{0,norm} < 0.1$), the TLS exchanges its population coherently with the qubit, as shown in Fig.~\ref{fig:tddata}(c).

When the strongest coupled TLS is not in the vicinity of the JJ, we found that the ensemble effect of many TLSs is pronounced. To demonstrate this result, we varied the number of strongly coupled TLSs included in the model from 1 to 200. 
We confirm that if the strongest coupled TLS is within $0.3\,\mathrm{\upmu m}$ from the center of the JJ, that TLS dominates the qubit dynamics, and the contribution from the rest of the coupled TLSs is negligible, as shown in Fig.~\ref{fig:tdanalysis}(f). On the other hand, when the strongest coupled TLS is farther than $0.3\,\mathrm{\upmu m}$ from the JJ, the qubit $T_1^q$ depends on the number of simulated TLSs. In Fig.~\ref{fig:tdanalysis}(e), $T_1^q$ is $\sim7825\,\upmu\mathrm{s}$ when there is only one strongest coupled TLS interacting with the qubit. However, when we include all 200 coupled TLSs, $T_1^q$ drops to $\sim250\,\upmu\mathrm{s}$. The $T_1^q$ stabilizes when the number of TLSs is above 150.

\begin{center}
\begin{figure}[t]
\includegraphics[scale=1.0]{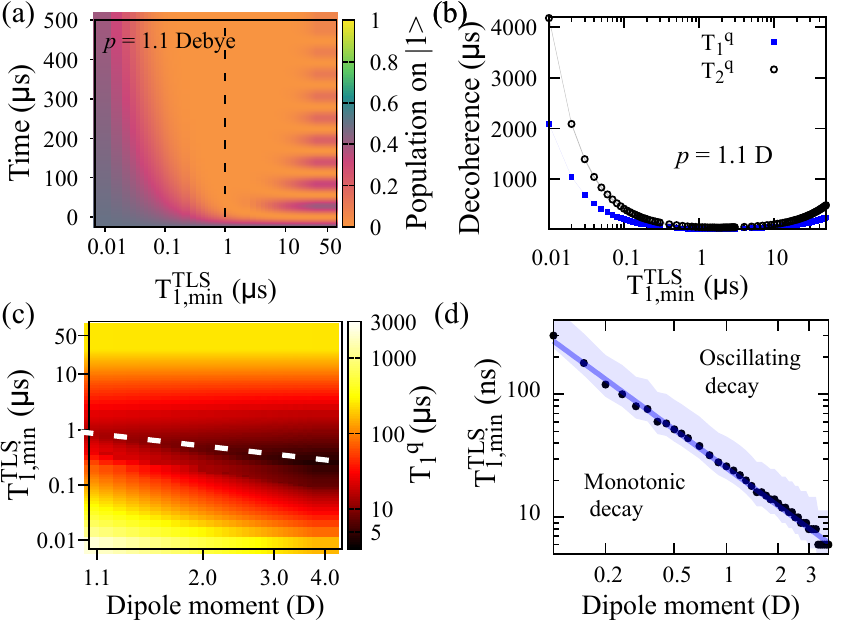}
\caption{Panel (a-c) show the change in the qubit dynamics when the strongest coupled TLS is $0.3-5\,\upmu\mathrm{m}$ away from the JJ. (a) Qubit dynamics change as a function of $T_{1,\mathrm{min}}^\textrm{TLS}$. The TLS dipole moment is fixed at 1.1 Debye. As $T_{1,\mathrm{min}}^\textrm{TLS}$ increases, we see the coherent exchange of qubit-TLS population. (b) $T_1^q$ (black) and $T_2^q$ (blue) times from the data in panel (a) show a minimum around $T_{1,\mathrm{min}}^\textrm{TLS}=1\,\upmu\mathrm{s}$ that increases as $T_{1,\mathrm{min}}^\textrm{TLS}$ moves away from $1\,\mathrm{\upmu s}$. (c) $T_1^q$ increases as the TLS dipole moment decreases and changes with $T_{1,\mathrm{min}}^\textrm{TLS}$ as in panel (b). The color indicates the $T_1^q$ time in log scale. (d) The black scatter points are the shortest $T_{1,min}^\textrm{TLS}$ time for the dipole moment between 0.1 and 4.0 Debye across multiple distributions of the TLSs. The shaded area indicates the standard deviation of the $T_{1,min}^\textrm{TLS}$ time.}
\label{fig:t1min}
\end{figure}
\end{center}

An adjustable parameter, $T_{1,\textrm{min}}^\textrm{TLS}$, provides insight into the microscopic origin of the TLSs. In principle, $T_{1,\textrm{min}}^\textrm{TLS}$ can be determined by the material properties of the TLSs, such as their energy, phonon velocity, and elastic field coupling strength \cite{phillips_two-level_1987}. However, estimating those parameters experimentally or theoretically  can result in large uncertainties. In this work, we use experimental observations, coupled with the insight gained from qubit-TLS simulations, to determine a reasonable range of values for $T_{1,\textrm{min}}^\textrm{TLS}$. 

Figure~\ref{fig:t1min}(a) shows the time-dependent qubit excited state population over $500\,\upmu\mathrm{s}$ at different $T_{1,\textrm{min}}^\textrm{TLS}$ with a fixed dipole moment of $1.1\,\textrm{Debye}$. When $T_{1,\textrm{min}}^\textrm{TLS}$ is below $1\,\upmu\mathrm{s}$, the qubit population in the excited state monotonically decays to the ground state. In this regime, the qubit $T_1^q$ time increases as $T_{1,\textrm{min}}^\textrm{TLS}$ decreases. Above $T_{1,\textrm{min}}^\textrm{TLS}=1\,\upmu\mathrm{s}$, oscillations start to appear due to coherent population exchange between the qubit and TLS defects. The black dashed line in panel (a) indicates $T_{1,\textrm{min}}^\textrm{TLS}$ when the oscillation starts to appear. 

The qubit relaxation times, $T_1^q$, obtained from Fig.~\ref{fig:t1min}(a), are shown in Fig.~\ref{fig:t1min}(b) with blue square points. We repeated those simulations with the initial state $(|0\rangle+|1\rangle)/2$ to obtain $T_2^q$ times. When $T_{1,\textrm{min}}^\textrm{TLS}$ is greater than $1\,\upmu\mathrm{s}$, both $T_1^q$ and $T_2^q$ are positively correlated with $T_{1,\textrm{min}}^\textrm{TLS}$. On the other hand, when $T_{1,\textrm{min}}^\textrm{TLS}$ decreases, the coherence times again increase due to the quantum Zeno effect \cite{misra_zenos_2008}, which says that a quantum state that is measured frequently does not evolve.

Figure~\ref{fig:t1min}(c) shows $T_1^q$ times with different TLS dipole moments ($x$-axis) and $T_{1,\textrm{min}}^\textrm{TLS}$ ($y$-axis). The $T_1^q$ times show the same trend as in panel (b) at different TLS dipole moments. As the TLS dipole moment becomes smaller, $T_1^q$ time increases due to the weaker potential energy of the TLS dipole at its local electric field, $\vec{p}\cdot\vec{E}$.

Here, we define {\emph{$T_{1,\textrm{min}}^\textrm{TLS}$ threshold}} as the $T_{1,\textrm{min}}^\textrm{TLS}$ when oscillations start to appear in the qubit excited state population and $T_1^q$ is the lowest. In experiments on transmon qubits, we rarely observe such oscillations of coherent population exchange between a qubit and TLS defects. This will be discussed more in the following section. From this experimental observation, we can determine the shortest $T_1$ time of TLS defects that is consistent with experimental results. The white dashed line in panel (c) denotes the $T_{1,\textrm{min}}^\textrm{TLS}$ when the oscillations start.

The $T_{1,\textrm{min}}^\textrm{TLS}$ threshold depends on the distribution of the TLSs. To find the shortest $T_{1,\textrm{min}}^\textrm{TLS}$ threshold, we repeated the simulation over 300 trial cases with dipole moments varying between 0.1 and 4.0 Debye, which are the expected values of the dipole moments in superconducting qubits \cite{lisenfeld_electric_2019, hung_probing_2022}. From the $T_{1,\textrm{min}}^\textrm{TLS}$ thresholds between 6 and 300 ns, the estimated upper bounds of $T_1^\textrm{TLS}$ are $10-460\,\mathrm{ns}$ in this range of dipole moments, in agreement with previously reported values \cite{lisenfeld_measuring_2010, lisenfeld_decoherence_2016, khalil_landau-zener_2014}.

\section{Discussion}
 In experiments, the qubit relaxation time, $T_1^q$, occasionally drops sharply when $T_1^q$ is monitored over a long period of time \cite{klimov_fluctuations_2018}. This is likely caused by thermal fluctuators changing the resonant TLS frequencies over time \cite{faoro_microscopic_2008, burnett_evidence_2014}. If a resonant TLS frequency shifts away from the qubit, the coupling becomes negligible. Conversely, if the TLS frequency gets closer to the qubit frequency, it becomes resonant with the qubit. This time dependence implies that the resonant TLSs are gradually resampled from the ensemble distribution of TLSs. As in Fig.~\ref{fig:tddata}, our statistical studies over many different TLS distributions on a representative chip geometry show us that the $T_1^q$ drops when the strongest coupled TLS is located in a region of strong electric fields, even if they are not located in the JJ. However, coupling strength does not guarantee the reduced $T_1^q$. The coupling strength depends on the tunneling energy and the dipole coupling strength, $\vec{p}\cdot\vec{E}$. 
Statistically, in Fig.~\ref{fig:tdanalysis}(d), the coupling strengths of the strongest coupled TLS in eachl trial case was found to be above $\sim2\,\mathrm{kHz}$. Due to the large increase in electric field close to the junction, when the $\vec{p}\cdot\vec{E}$ for the strongest coupled TLS is high, a wide range of $\Delta_{0,norm}$ values can be the strongest coupled TLS in each trial. Therefore, $\Delta_{0,norm}$ alone cannot be a predictor of $T_1^q$. On the other hand, if the $\vec{p}\cdot\vec{E}$ is low, $\Delta_{0,norm}$ must be high to remain as the strongest coupled TLS. Interestingly, in the upper left corner of Fig.~\ref{fig:tdanalysis}(d), where  $T_1^q$ is long, $T_1^{TLS}$ is short due to high $\Delta_{0,norm}$ from Eq.(\ref{eqn:T1min}) and the accumulated population in the TLS excited states is low, which supports the data in Fig.~\ref{fig:tddata}(b), inset. A fruitful direction of a future study would be to understand how a short $T_1^{TLS}$ leads to a long $T_1^q$.

In most of the trials, we found that the electric field of the strongest coupled TLS is not sufficient to significantly reduce the $T_1^q$ time. In these cases, $T_1^q$ highly depends on the number of TLSs we include in the simulation, decreasing the $T_1^q$ by a factor of 30 as we increase the number of simulated TLS from 1 to 200. Our simulations suggest that in a real experiment, there are at least 150 TLSs participating in the qubit relaxation processes unless there is a $T_1^q$ drop-out.

One of the important adjustable parameters in our simulation is $T_{1,\textrm{min}}^\textrm{TLS}$, the shortest TLS $T_1$ time. 
In our model, when we increase $T_{1,\textrm{min}}^\textrm{TLS}$, oscillations between the qubit and the TLS excited states are observed in all cases. 
Though theoretically expected, such oscillations have not been experimentally observed in transmon qubits unless a strongly coupled TLS is precisely tuned to be in resonance with the qubit \cite{lisenfeld_electric_2019}.  
For the $T_{1,\textrm{min}}^\textrm{TLS}$ values chosen to be on the same order of magnitude as experimental results, the qubit energy relaxes monotonically in $\sim98.5\%$ of cases, while the remaining $\sim1.5\,\%$ shows oscillations between the qubit and the total TLS population. In this study, we specifically chose a typical transmon geometry.
If the Josephson junction area is large $(>1\,\upmu\mathrm{m}^2)$, such as in phase qubits, the coupling could be much stronger \cite{grabovskij_strain_2012} and oscillations can be observed more frequently \cite{lisenfeld_observation_2015}. Since we are in the regime where the oscillations are rarely observed, we can put limits on $T_{1,\textrm{min}}^\textrm{TLS}$. We estimate the $T_{1,\textrm{min}}^\textrm{TLS}$ is at most $6-300\,\mathrm{ns}$ for the dipole moment between $1.1$ and $4.0\,\mathrm{Debye}$.

In this model, we find that $T_2^q$ is exactly twice of the $T_1^q$ time, meaning that resonant TLSs alone do not affect the dephasing. If thermal fluctuators with energies that are comparable to the environmental thermal energy ($\sim 10\,\mathrm{mK}$) are included, they introduce slowly-varying TLS frequencies, leading to qubit dephasing. In future studies, including thermal fluctuators and observing ensemble effects of TLSs to study spectral diffusion would be fruitful.

\section{Conclusion}

This study reveals the complex processes occurring during qubit relaxation in an open quantum system. 
In this work, we simulate how ensembles of resonant defects change qubit energy relaxation times on a realistic quantum processor design. As expected, we find that strongly coupled TLSs are primarily located in strong electric field regions, such as at the edges of metal structures and in the Josephson junction. We find that it is actually the position of the strongest coupled TLS that is correlated with the qubit dynamics, instead of its coupling strength. When the strongest coupled TLS is in the vicinity of the JJ, it dominates the qubit dynamics. On the other hand, if the strongest coupled TLS is located further away from the JJ, the qubit dynamics require at least 150 TLSs for the simulated qubit lifetime, $T_1^q$, to converge to an asymptotic value. From our simulation results, we are able to infer bounds on the decay times of TLS that can elucidate the microscopic origin of decoherence inducing defects.

This work bridges the existing gap between single TLS-qubit models and a realistic picture of many TLS affecting a qubit in the STM. The framework that was developed in this work can provide guidance for optimizing future quantum processor designs for longer coherence times and less sensitivity to defects.

\section{Acknowledgement}
This work was primarily supported by LDRD 20-ERD-010 (all aspects until September 2022). This work was also supported by the U.S. Department of Energy, Office of Science, Basic Energy Sciences, under Award DE-SC0020313 (manuscript preparation from October 2022 onwards). Additionally, this work was performed under the auspices of the U.S. Department of Energy by Lawrence Livermore National Laboratory under Contract DE-AC52-07NA27344. D. J. acknowledges the support of the Computational Science Graduate Fellowship from the U.S. Department of Energy under Grant No. DE-SC0019323. 

\section{Data Availability}
The data that support the findings of this study are available upon request from the authors.


%

\end{document}